\documentclass[11pt,twoside]{article}

\usepackage{asp2006}
\usepackage{epsf}
\usepackage{psfig}
\usepackage{lscape}

\begin{document}
\title{Detecting Obscured AGN in the Distant Universe with \textit{Spitzer}} 
\author{
J. L. Donley, \altaffilmark{1} G. H. Rieke, \altaffilmark{1}
P. G. P\'{e}rez-Gonz\'{a}lez, \altaffilmark{1,2} J. R. Rigby,
\altaffilmark{1} A. Alonso-Herrero \altaffilmark{3}}

\altaffiltext{1}{Steward Observatory, University of Arizona, 933 
North Cherry Avenue, Tucson, AZ 85721; jdonley@as.arizona.edu}
\altaffiltext{2}{Departamento de Astrof\'{\i}sica y CC. de la Atm\'osfera, Facultad de
CC. F\'{\i}sicas, Universidad Complutense de Madrid, 28040 Madrid,
Spain}
\altaffiltext{3}{Departamento de Astrof\'{\i}sica Molecular e Infrarroja, 
Instituto de Estructura de la Materia, CSIC, E-28006 Madrid, Spain}

\begin{abstract} 

We present the results of a \textit{Spitzer} search for obscured AGN
in the \textit{Chandra} Deep Field-North, using both radio-excess and
mid-infrared power-law selection. AGN selected via the former
technique tend to lie at $z \sim 1$, have SEDs dominated by the 1.6
\micron\ stellar bump, and have Seyfert-like X-ray luminosities (when
detected in the X-ray).  In contrast, the IRAC (3.6-8.0 \micron)
power-law selected AGN lie at higher redshifts of $z \sim 2$, and
comprise a significant fraction of the most X-ray luminous AGN in the
CDF-N.  While there is almost no overlap in the AGN samples selected
via these two methods, their X-ray detection fractions are very
similar.  Only 40\% and 55\% of the radio-excess and power-law samples
are detected in the 2~Ms X-ray catalog, respectively.  The majority of
the AGN selected via both methods are consistent with being obscured
($N_{\rm H} > 10^{22}$~cm$^{-2}$), but not Compton-thick ($N_{\rm H} >
10^{24}$~cm$^{-2}$), although Compton-thick candidates exist in both
samples.  We place an upper limit of $\le 82$\% (or $\le 4:1$) on the
obscured fraction of the power-law sample, consistent with predictions
from the cosmic X-ray background.  The sources selected via the
power-law criteria comprise a subset of AGN selected via other IRAC
color-color cuts. While smaller in number than the color-selected
samples in the deep fields, the power-law sample suffers from less
contamination by star-forming galaxies.

\end{abstract}

\section{Introduction}

Defining complete samples of AGN, both locally and in the distant
cosmological fields, has been a major ongoing goal of AGN research.
Hard X-ray selection is generally considered the best way to detect
both relatively uncontaminated and complete samples of AGN.  At high
column densities, however, the dust and gas surrounding the central
engine (in combination with that located in the host galaxy) are
capable of hiding virtually all accessible AGN tracers. Therefore,
while the overall resolved fraction of the cosmic X-ray background
(CXRB) is high, it drops with increasing energy to 60\% at 6-8~keV and
to 50\% at $>$ 8 keV
\citep{2004MNRAS.352L..28W,2005MNRAS.357.1281W}. Population synthesis
models of the CXRB therefore predict a significant population of
heavily obscured AGN not detected in the deepest X-ray fields.  For
instance, \citet{2004ApJ...616..123T} predict that current X-ray
catalogs are 25\% incomplete at $N_{\rm H} = 10^{23}$~cm$^{-2}$ and
70\% incomplete at $N_{\rm H} = 10^{24}$~cm$^{-2}$.

Numerous attempts have been made to detect this population of heavily
obscured AGN, many of which have focused on the mid-infrared (MIR)
emission where the obscured radiation is re-emitted or on combinations
of MIR and multi-wavelength data. We focus here on two selection
techniques independent of the optical and X-ray properties of the AGN:
radio-excess and MIR power-law selection.  The former selects AGN
whose radio emission far exceeds that of the radio-infrared
correlation for star-forming galaxies, whereas the latter selects AGN
whose IRAC SEDs exhibit the characteristic power-law emission expected
of luminous AGN \citep[e.g.][]{1994ApJS...95....1E}. By focusing on the
infrared and radio, wavebands where dust obscuration is minimal, we
can select both unobscured AGN as well as heavily obscured AGN likely
to be missed in the UV, optical, and soft X-ray bands.

To test the X-ray properties of the \textit{Spitzer}-selected AGN
samples, we focus on the CDF-N, the deepest X-ray field observed to
date.  \textit{Spitzer} MIPS and IRAC images with exposures of 1400~s
and 500~s, respectively, cover the full area of the \textit{Chandra}
2~Ms field, as does the deep 1.4 GHz radio data of
\citet{2000ApJ...533..611R}.  Optical and near-infrared imaging and
photometry are available from the GOODS dataset
\citep{2004ApJ...600L..93G} as well as from the data of
\citet{2004AJ....127..180C} (\textit{UBVRIz'HK'}). We assume a 
cosmology of ($\Omega_{\rm m}$,$\Omega_{\rm \Lambda},H_0$)=(0.3, 0.7,
72~km~s$^{-1}$~Mpc$^{-1}$).

\section{Radio-excess Sample}

We use the well-known radio-infrared relation of star-forming galaxies
and radio-quiet AGN, to select a sample of radio-excess
AGN. \citet{2004ApJS..154..147A} define $q$ = log
(f$_{24~\micron}$/f$_{1.4~\rm GHz}$). For star-forming galaxies, $q$
is tightly constrained out to z $\sim$ 1$:$ $q$ = 0.94 $\pm$ 0.23
after K-correction \citep{2004ApJS..154..147A}. Galaxies with values
of $q$ well below this range (strong radio with respect to 24~\micron\
emission) are unlikely to be dominated by star formation and instead
are radio-emitting AGN \citep[e.g.][]{1995AJ....109.2318C}.  We
therefore set a selection threshold of $q < 0$ to classify a galaxy as
probably having an AGN and require a \textit{Chandra} exposure of $>$
1~Ms to provide high X-ray sensitivity. This selection identifies 27
radio-excess AGN in the CDFN
\citep{2005ApJ...634..169D}.

The radio-excess AGN lie at the typical redshift of X-ray sources in
the CDF-N, $z \sim 1$, and (when detected in the X-ray) tend to have
X-ray luminosities typical of Seyfert galaxies and SEDs dominated by
the 1.6 \micron\ stellar bump.  Approximately 60\% of these
radio-excess AGN, however, are X-ray undetected in the 2~Ms
\textit{Chandra} catalog, even at exposures of $\ge$ 1~Ms; 25\% lack
even 2$\sigma$ X-ray detections. The absorbing columns to the faint
X-ray-detected objects are $10^{22}$ cm$^{-2} <$ N$_H$ $< 10^{24}$
cm$^{-2}$, i.e., they are obscured but unlikely to be Compton
thick. Of the 9 AGN that are either X-ray non-detected or are too
close to a known X-ray source to search for weak X-ray emission, we
estimate that 6 could be Compton thick.  This corresponds to 22\% of
our sample and is consistent with predictions from the X-ray
background.

\section{MIR Power-law Sample}

In the MIR, luminous AGN can often be distinguished by their
characteristic power-law emission, which extends from the infrared to
the ultraviolet \citep[e.g.][]{1979ApJ...230...79N,1994ApJS...95....1E}.
This emission is not necessarily due to a single source, but can arise
from the combination of non-thermal nuclear emission and thermal
emission from various nuclear dust components
\citep[e.g.][]{1981ApJ...250...87R}.

\citet{2006ApJ...640..167A} selected a sample of 92 such sources in
the CDF-S, 70\% of which are hyper-luminous infrared galaxies
(HyperLIRGS, log L$_{\rm IR}$(L$_{\sun}$)$ > 13$) or ultra-luminous
infrared galaxies (ULIRGs, log~L$_{\rm IR}$(L$_{\sun}$)$ >
12$). Nearly half (47\%) of their power-law sample were undetected in
X-rays at exposures of up to 1~Ms. We use a selection similar to that
of \citet{2006ApJ...640..167A} to identify a sample of 62 high $S/N$,
red ($\alpha \le -0.5$, where $f_{\nu} \propto \nu^{\alpha}$) IRAC
power-law galaxies in the 2~Ms CDF-N \citep{donley07}.

\subsection{X-ray Properties}

Of the 62 power-law galaxies, only 34 (55\%) have X-ray counterparts
in the 2~Ms CDF-N X-ray catalog \citep{2003AJ....126..539A},
consistent with the findings of \citet{2006ApJ...640..167A}, although
85\% show evidence for X-ray emission at the $\ge 2.5 \sigma$
detection level.  While the power-law galaxies comprise only $\sim
20$\% of the X-ray and MIR-detected AGN in the CDF-N, they make up a
significant fraction of the high redshift ($z \sim 2$) and high
luminosity X-ray AGN sample, as is demonstrated in Fig. 1. This is not
surprising, as the power-law selection requires the AGN to be
energetically dominant, and therefore preferentially selects the most
luminous AGN.  To illustrate the effect of X-ray luminosity on the AGN
contribution to the optical through MIR continuum, we also plot in
Fig. 1 the median rest-frame optical-MIR SEDs of the X-ray--detected
members of the comparison sample, as a function of X-ray
luminosity. The comparison sample consists of the 1420 IRAC sources in
the CDF-N that meet our $S/N$ and X-ray exposure cuts. Low-luminosity
X-ray sources are dominated by the stellar bump in the optical-NIR
bands \citep[e.g.][]{2004ApJS..154..155A}.  The strength of this
feature decreases with increasing X-ray luminosity, and disappears
almost entirely at luminosities of log~$L_{\rm x}$(ergs~s$^{-1}$)$ >
44$, where the SED takes on the characteristic power-law shape.

\begin{figure}[ht!]
\centering
\plottwo{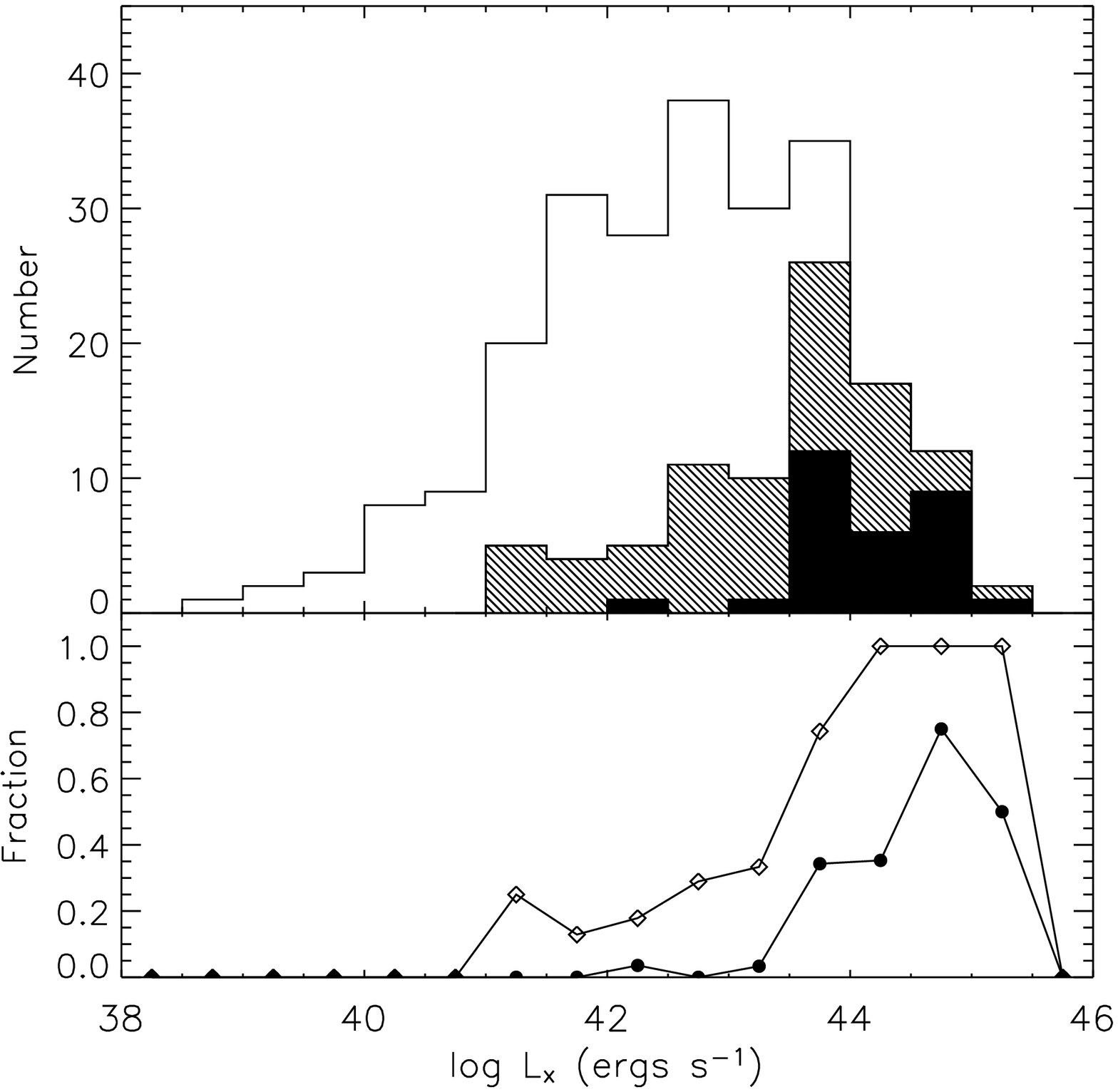}{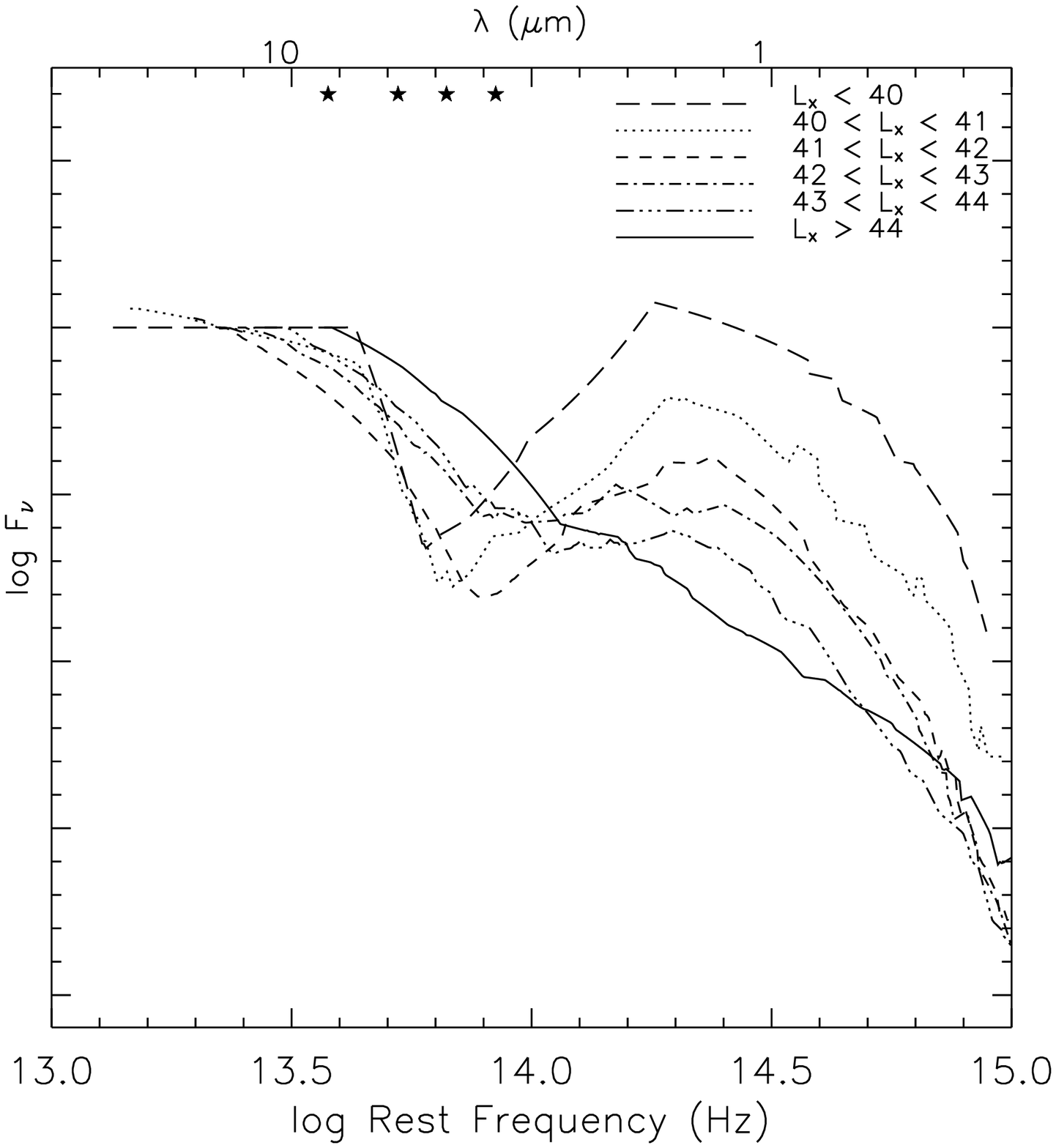}
\caption{Left: X-ray luminosity distributions of X-ray--detected 
IRAC sources (unshaded histogram), the power-law sample (filled
histogram), and the sample of AGN selected by the
\citet{2004ApJS..154..166L} color criteria (lightly shaded). The lower
panel gives the fraction of the X-ray sources that meet the power-law
(filled circles) or Lacy et al. (open diamonds) criteria. Right:
Median rest-frame SEDs of the X-ray--detected IRAC sources, as a
function of observed 0.5-8 keV luminosity (in units of
ergs~s$^{-1}$). Stars indicate the wavelengths of the four IRAC
bands. }
\end{figure}

\subsection{Obscuration}

The obscured fraction ($N_{\rm H} > 10^{22}$~cm$^{-2}$) of the
X-ray--detected power-law galaxies is $\sim$~68\%, in agreement with
that of \citet{2003ApJ...598..886U} for AGN at similar redshifts.  If
we consider only those power-law galaxies in the X-ray catalog, we
therefore derive an obscured to unobscured ratio of $\sim$~2:1.
Adding the 10 weakly-detected power-law galaxies, all of which are
likely to be obscured, we calculate an obscured fraction of
$\sim$~75\%.  If we further include the power-law galaxies not
detected in X-rays, assuming that all are obscured, the maximum
obscured ratio of power-law galaxies rises to $<$ 4:1-5:1 (82\%).

Power-law galaxies detected to high $S/N$ in the IRAC bands account
for at most $\sim 20-30$\% of both the MIR-detected AGN and the
MIR-detected obscured AGN predicted by the X-ray luminosity function
synthesis models of \citet{2006ApJ...640..603T} down to 24 \micron\
flux densities of $80$ $\mu$Jy. The majority of obscured AGN should
therefore have SEDs dominated by the host galaxy, or red power-law
SEDs that fall below the IRAC detection limit.

\subsection{Comparison to IRAC Color-Selection}

We plot in Fig. 2 the position of the power-law galaxies with respect
to the IRAC AGN color cuts of \citet{2004ApJS..154..166L} and
\citet{2005ApJ...631..163S}, designed for relatively shallow surveys.
All of the power-law galaxies lie within both the Lacy et al. and
Stern et al.  color regions, and as such, they comprise a subset of
the color-selected MIR sources.  While the color-selected AGN samples
in the CDF-N comprise a higher fraction of high-luminosity AGN than
the power-law selected sample (see Fig. 1), they also appear to suffer
from a higher degree of contamination from star-forming galaxies (see
Table 1).  This is not surprising, given the behavior of the
star-forming and AGN templates shown in Fig. 2 \citep[available
in][]{donley07}.  The star-forming templates enter the AGN selection
regions at both low and high redshifts, whereas they fall outside of
the more stringent power-law region at $z<2.8$.

\begin{figure}[ht!]
\centering
\plottwo{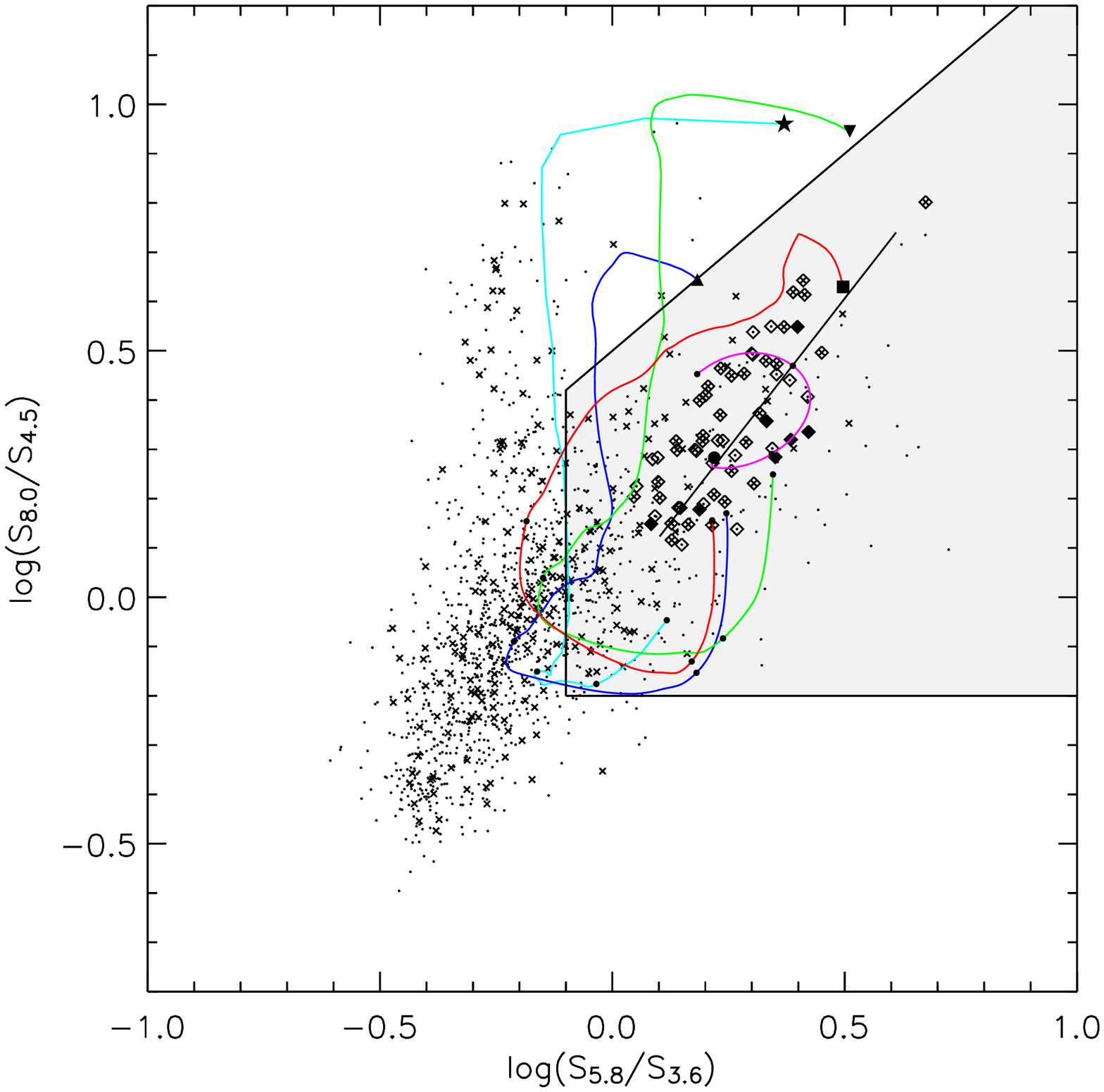}{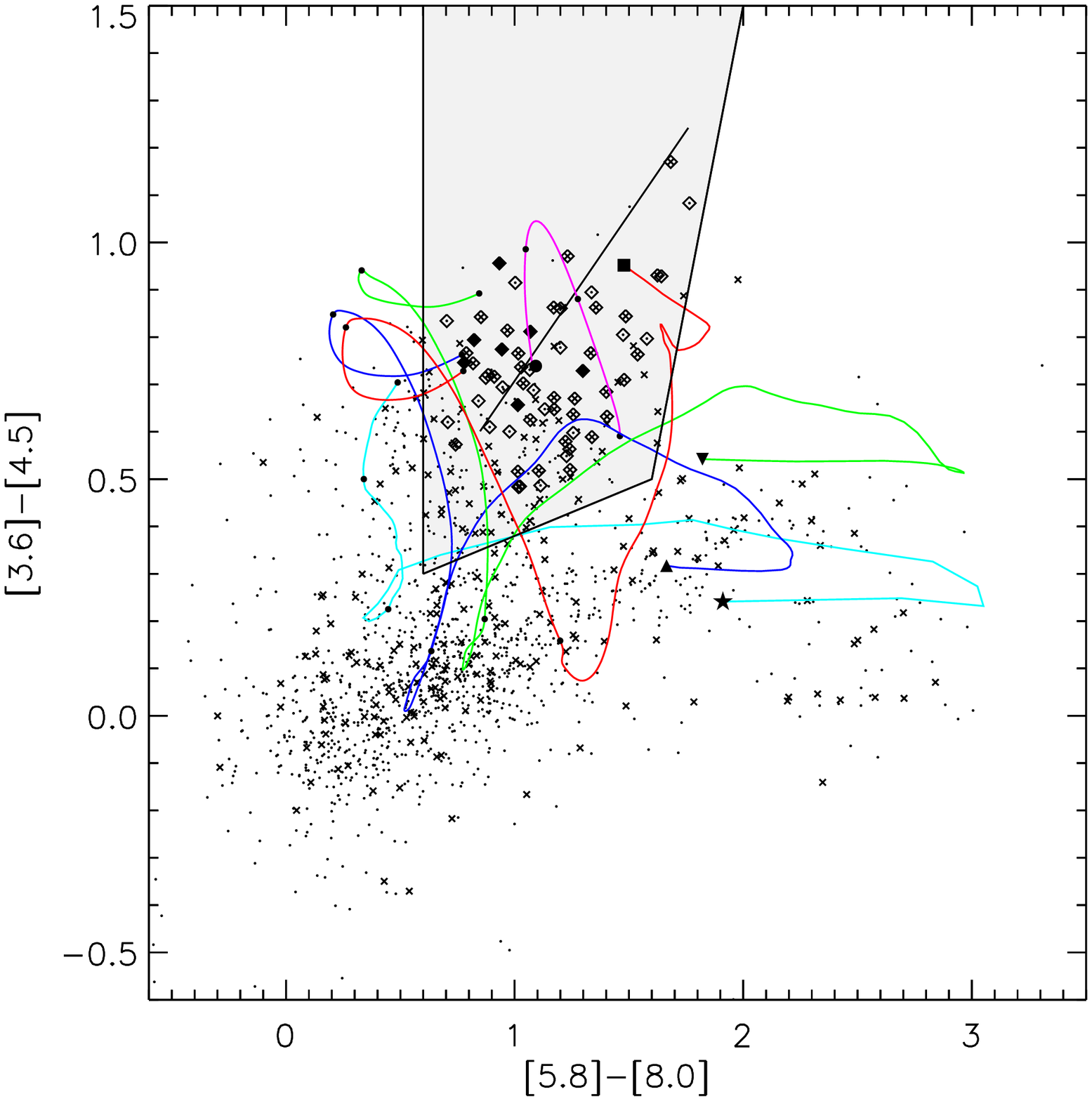}
\caption{Location of CDF-N IRAC sources on the color-color diagram
of \citeauthor{2004ApJS..154..166L} (2004, left) and
\citeauthor{2005ApJ...631..163S} (2005, right). Power-law galaxies are
given by diamonds and X-ray sources are given by crosses. The 7
power-law galaxies that are X-ray non-detected to the 2.5 $\sigma$
level are shown as filled diamonds. Overplotted are the redshifted
IRAC colors of a typical star-forming galaxy
\citep[cyan star]{2002ApJ...576..159D}, the starburst-dominated ULIRGs
Arp 220 (blue triangle) and IRAS~17208 (green upside down triangle),
the ULIRG/Sey 2 Mrk 273 (red square) and the radio-quiet AGN SED from
\citeauthor{1994ApJS...95....1E} (1994, magenta circle), where the
indicated point represents the colors at $z=0$, and small circles mark
the colors at $z=1,2,\rm{ and}\ 3$. The star-forming templates enter
the color-selection regions at both low and high redshifts, but fall
outside of the more stringent power-law region at $z < 2.8$.}
\end{figure}

\begin{table}[!ht]
\caption{Properties of power-law and color-selected CDF-N samples}
\smallskip
\begin{center}
{\small
\begin{tabular}{lccc}
\tableline
\noalign{\smallskip}
Selection Criteria  & Power-law  & Stern et al. &  Lacy et al. \\
\noalign{\smallskip}
\tableline
\noalign{\smallskip}
Number                              & 62\ \ \ & 216\ \  & 393     \\
Detected in X-rays                  & 55\%  & 38\%  &25\% \\
Detected at 24\micron               & $>$92\%  & 83\% & 72\% \\
L$_{\rm 1.4 GHz} > 10^{24}$~W~Hz$^{-1}$ & 88\% & 54\%  & 49\% \\
Quiescent Optical/X-ray Colors      & $\ge 3\%$ & $\ge 10\%$ & $\ge 15\%$ \\ 
\noalign{\smallskip}
\tableline
\end{tabular}
}
\end{center}
\end{table}

\section{Summary}

We present the results of a \textit{Spitzer} search for AGN in the
\textit{Chandra} Deep Field-North.  We focus on two selection
techniques independent of the optical and X-ray properties of the
sources.  The first is radio-excess selection, in which we select
sources whose radio emission far exceeds that of the radio-infrared
correlation.  In the CDF-N, we detect 27 radio-excess AGN.  These
sources tend to lie at $z \sim 1$ and have SEDs dominated by the 1.6
\micron\ stellar bump.  Only 40\% are detected in the X-ray at
exposures of $> 1$~Ms, although 75\% show signs of X-ray emission at
the $>2\sigma$ detection level. Those with X-ray counterparts have
Seyfert-like X-ray luminosities.  The majority of the radio-excess AGN
are consistent with being obscured ($N_{\rm H} > 10^{22}$~cm$^{-2}$),
and 20\% are potentially Compton-thick.

The second selection technique is MIR power-law selection, in which we
select sources with red ($\alpha < -0.5$) IRAC power-law SEDs.  The
sources selected in this way tend have higher redshifts ($z \sim 2$)
and X-ray luminosities than the radio-excess sample.  The IRAC
power-law selection recovers a significant fraction of the most X-ray
luminous AGN in the CDF-N.  Despite this, only 55\% of the power-law
galaxies are detected in the X-ray, with 15\% remaining undetected
down to the $>2.5\sigma$ detection level. These X-ray detection
fractions are similar to those of the radio-excess AGN, even though
there is almost no overlap in the two samples. This suggests that a
relatively low X-ray detection rate may be a common feature for AGN
samples selected independently of their optical and X-ray properties,
regardless of luminosity or redshift.

A study of the intrinsic obscuration of the power-law galaxies
suggests that as many as 82\% (4:1 - 5:1) are obscured. The number
densities of these sources indicate that the power-law sample
comprises only 20-30\% of the MIPS-detected obscured AGN population.
The remaining AGN are likely to have SEDs dominated by the host
galaxy, like those of the radio-excess sample.

The power-law galaxies discussed here comprise a subset of AGN
selected via the IRAC color-selection criteria of
\citet{2004ApJS..154..166L} and \citet{2005ApJ...631..163S}.  While
larger in number than the power-law sample, the color-selected samples
in the CDF-N appear to suffer from a higher degree of contamination
from star-forming galaxies.  By applying additional SED constraints to
a color-selected sample, or by requiring either a 24 \micron\
detection or AGN-like optical/X-ray colors, one can create both a more
complete and reliable sample of MIR-selected AGN, even in the deepest
astronomical fields.  Doing so is crucial if we are to constrain the
fraction of heavily obscured AGN in the distant universe and determine
their contribution to the accretion history of the universe.

\acknowledgements 

This work was supported by an NSF Graduate Research Fellowship and by
NASA through contracts 960785 and 1255094, issued by JPL/California
Institute of Technology.

\end{document}